\documentclass[epsfig,12pt]{article}
\usepackage{epsfig}
\def\beqn{\begin{eqnarray}}
\def\eeqn{\end{eqnarray}}

\def\beq{\begin{equation}}
\def\eeq{\end{equation}}
\def\ba{\beq\new\begin{array}{c}}
\def\ea{\end{array}\eeq}

\newcommand{\ntwo}{${\mathcal N}=2\;$}
\newcommand{\none}{${\mathcal N}=1\;$}

\newcommand{\pt}{\partial}

\renewcommand{\theequation}{\thesection.\arabic{equation}}

\begin{document}

%%%%%%%%%%%%%%%%%%%%%%%%%%%%%%%%
%%%%%%%%%%%%%%%%%%%%%%%%%%%%%%%%

\begin{titlepage}

\begin{flushright}
FTPI-MINN-07/18, UMN-TH-2605/07\\
ITEP-TH-20/07\\
\end{flushright}

\vspace{1cm}

\begin{center}

{\Large \bf
Confinement in \none SQCD:\\[2mm]
One Step Beyond Seiberg's Duality
 }
\end{center}

\vspace{3mm}

\begin{center}
{\bf M.~Shifman$^{a}$ and \bf A.~Yung$^{a,b,c}$}
\end {center}
\vspace{0.3cm}
\begin{center}

$^a${\it  William I. Fine Theoretical Physics Institute,
University of Minnesota,
Minneapolis, MN 55455, USA}\\
$^{b}${\it Petersburg Nuclear Physics Institute, Gatchina, St. Petersburg
188300, Russia}
$^c${\it Institute of Theoretical and Experimental Physics, Moscow
117259, Russia}\\

\end{center}

\begin{abstract}

We consider \none supersymmetric quantum chromodynamics
(SQCD)  with the gauge group U($N_c$) and
$N_c+N$ quark flavors. $N_c$ flavors are
massless; the corresponding squark fields develop (small) vacuum expectation values
(VEVs) on the Higgs branch. Extra $N$ flavors are endowed with
small (and equal) mass terms. We study this theory through its Seiberg's dual:
U$(N)$ gauge theory with $N_c +N$ flavors of ``dual quark" fields plus a
gauge-singlet mesonic field $M$.
The original theory is referred to as ``quark theory"
while the dual one is termed ``monopole theory."
The suggested mild deformation
of Seiberg's procedure changes the dynamical regime of the monopole theory
from infrared free to asymptotically free at large distances.
We show that, upon condensation of
the ``dual quarks," the dual theory supports
non-Abelian flux tubes (strings).
Seiberg's duality is extended beyond purely massless states to include
light states on both sides.
Being interpreted in terms
of the quark theory, the monopole-theory flux tubes are supposed to carry chromoelectric fields.
The string junctions ---  confined monopole-theory monopoles
--- can be viewed as ``constituent quarks" of the original quark theory.
We interpret closed strings as glueballs of the original quark theory.
Moreover, there are
string configurations formed by two junctions connected by a pair of
different non-Abelian strings. These can be considered as constituent
quark mesons of the quark theory.

\end{abstract}

\end{titlepage}

\newpage

\section{Introduction}
\label{intro}
\setcounter{equation}{0}

In the mid-1990s Seiberg argued \cite {Sdual,Seiberg1} that two distinct
\none Yang--Mills theories with appropriately chosen matter
(usually referred to as electric/magnetic theories)
 can be equivalent in the infrared (IR) limit.
Seiberg's duality was tested in a number of nontrivial ways
(it passed all tests with flying colors), was understood in the framework of
string/D brane theory and became an important tool in the realm of strongly
coupled gauge theories. Seiberg's duals have different gauge groups,
say, SU$(N_c)$ in the electric theory versus
SU$(N_f-N_c)$ in the magnetic one.
The matter sectors of the electric and magnetic theories from the dual pair
are related to each other in a well-defined manner.

Seiberg's duality was most extensively used in the so-called conformal window \cite{Shifman1Rev}, in which both the original theory and its dual are conformally invariant.
Typically, while one theory is strongly coupled the other one is at weak coupling
which allows one to perform fully quantitative analysis of the weakly coupled
theory, with predictions for scaling dimensions
applicable to the strongly coupled dual theory in the
conformal regime.

Seiberg's duality holds outside the conformal window too.
It relates to each other massless states of the dual
theories, one of which is infrared free. Moreover, by itself it does not provide us with a confinement mechanism
based on formation of confining strings (flux tubes).
The questions we address are:
can Seiberg's duality be generalized to include theories which at large distances exhibit
non-Abelian string-based confinement?
Can this duality connect with each other not only massless states but also massive ones? 

Below we will argue that the answers to these two  questions 
are positive. To construct dual pairs with the above properties we will need a modest extension of ideas developed in connection with Seiberg's duality previously
\cite{Shifman1Rev,IS,ISS}. As well-known \cite{Sdual,Seiberg1,Shifman1Rev,IS},
large vacuum expectation values (VEVs) of the squark fields in the electric theory
translate into large  mass terms of the ``dual quarks"
in the magnetic theory and {\em vice versa}.
With large VEVs we Higgs a part of the gauge group.
The corresponding gauge bosons become heavy and can be integrated out.
On the other side of duality the corresponding ``dual quarks" become
heavy and can be integrated out. This procedure changes the IR behavior of each
theory from the given dual pair, but their IR equivalence remains intact.

We want to change the conformal (or IR free regime) in the infrared limit
into a confining regime. To this end the above procedure must be modified.
The modification we suggest is as follows.
We start from an \none SQCD with a certain number of
the quark fields. The dynamical scale of this theory is $\Lambda_Q$.
Quark fields from a judiciously chosen subset are assumed to develop VEVs which are
small on the scale of $\Lambda_Q$. The remaining quark fields are endowed with a
common mass term $m_q$ which is also small
compared to $\Lambda_Q$, so that all  ``hadrons" are dynamical;
none can be integrated out. We argue that within the framework
of this deformation of Seiberg's procedure, on the other side of duality,
the IR free regime is deformed to give rise to
a theory which supports flux tubes (strings) at weak coupling
and confines non-Abelian (dual) monopoles.
A number of states in this theory are light in the sense that
their mass tends to zero in  the limit $m_q\to 0$. We argue that via duality these light states
are in one-to-one correspondence with the light states of the original theory.
Thus, duality gets extended to include (in addition to massless moduli)
a part of the spectrum which is light
compared to the natural dynamical scale $\Lambda_Q$ {\em but not massless}.
We refer to such dual pairs as quark/monopole theories to distinguish
them from Seiberg's electric/magnetic theories.

Extended duality allows us to analyze the monopole theory at weak coupling
and make a number of highly nontrivial predictions for the quark theory light sector which is at strong coupling. In fact, the monopole theory we get is close to the so-called
$M$ model developed previously \cite{Mmodel} with the purpose
of studying non-Abelian strings and  confined non-Abelian monopoles in \none super-Yang--Mills
theories. Some distinctions of the monopole theory from the
$M$ model (e.g. non-BPS nature of the flux tubes) do not produce drastic changes
of the dynamical pattern we arrive at compared to the dynamical pattern of
the $M$ model.

As was mentioned, the
non-Abelian monopoles must be important in Seiberg's duality
being related  to ``dual quarks.'' We make one step further
suggesting that the
non-Abelian monopoles {\em are} the ``dual quarks.''
The dual quark fields condense prividing (small) masses
to all gauge bosons of the monopole theory.
The way the monopole theory is Higgsed is very peculiar ---
it corresponds to baryon-operator dominated vacuum in the  quark theory.
Confined monopoles of the monopole theory are to be interpreted as certain
``constituent quarks" of the quark theory. Both form $N$-plets of the
global unbroken SU$(N)$ theory which is present in the quark and monopole theories,
on both sides of the extended duality.

To explain how this works
we have to be more specific.
What are the non-Abelian monopoles?

Although the full answer is not yet known there are certain results which were
obtained recently and which
have a natural interpretation in the framework of ``the non-Abelian monopole"
hypothesis.

Let us start from the
Abelian 't Hooft-Polyakov monopole \cite{thopo} of which we know everything. Suppose we
have a model with the SU(2) gauge group broken down to U(1) by condensation of adjoint scalars
$a^a$, $a=1,2,3$. One can always align the adjoint condensate along the third axis
in the SU(2) space, $$\langle a^a \rangle=\delta^{a3}\langle a^3 \rangle\,.$$
The size of the  't Hooft-Polyakov monopole is of
the order of $\langle a^3 \rangle^{-1}$ while its mass is of the
order of $\langle a^3 \rangle/g^2$, where $g^2$ is the gauge coupling constant.
In the non-Abelian limit $\langle a^3 \rangle\to 0$ the size of this state
formally  tends to
infinity and its mass tends to zero, see \cite{We} for a thorough
discussion. The monopole
seems to disappear, at least classically, from the physical spectrum.

However, from recent supersymmetric studies we learned
that in some cases the monopoles become stabilized in the non-Abelian limit
by quantum effects. In particular, this happens to confined monopoles
in  \ntwo SQCD \cite{SYmon,HT2} with the gauge group U($N$), the Fayet-Iliopoulos (FI) term for the U(1) factor and
$N_f=N$ quark flavors. This theory supports non-Abelian flux tubes ---
strings \cite{HT1,ABEKY,SYmon,HT2}. They
are formed upon quark condensation triggered by the FI term. The string orientational
zero modes are associated with the rotation of their color
flux inside the non-Abelian gauge
subgroup SU($N$). Non-Abelian strings originate from the
Abelian $Z_N$ strings in the
special regime which ensures vanishing of VEVs of the adjoint scalars, i.e. exactly
in the non-Abelian limit we are interested in. The internal dynamics of the
orientational modes of the non-Abelian strings are described by two-dimensional
 (2,2) supersymmetric CP$(N-1)$ model on the string worldsheet
\cite{HT1,ABEKY,SYmon,HT2}.

The monopoles are confined by these stings. In fact, elementary monopoles are
nothing but the junctions
of elementary $Z_N$ strings \cite{Tong,SYmon}. It is possible to trace the fate of the
confined monopoles \cite{SYmon} --- from the Abelian regime (with nonvanishing VEVs of the  adjoint fields)
where they are just the 't Hooft-Polyakov monopoles slightly deformed by confinement effects,
deep into the non-Abelian limit of vanishing VEVs of the adjoint fields.
In this limit confined monopoles are seen as kinks of the CP$(N-1)$ model on the string
worldsheet. They are stabilized by quantum effects. Their size and mass are determined
 by the dynamical scale of the CP$(N-1)$ model. Thus, in quantum theory
 confined monopoles do not disappear from the physical spectrum in the
non-Abelian limit.\footnote{
A somewhat different approach to
non-Abelian monopoles was developed in \cite{Konishi}.}

As was recently shown \cite{Mmodel},  breaking
\ntwo supersymmetry down to \none we do not necessarily
destroy the above confined non-Abelian monopoles. In particular, they still exist in the limit
when the adjoint fields decouple  altogether
and no traces of ``Abelization'' of the theory remain.

These results lead us to the following  conjecture. Since the
monopoles (stabilized by quantum effects) survive in the non-Abelian limit
 in the confining phase it is plausible to suggest that
they can exist also in other phases matching Seiberg's notion of ``dual quarks."
It is natural then that the theory  dual to \none SQCD is formulated as a gauge theory
of non-Abelian monopoles. In particular, in our construction the monopoles
(a.k.a the quark fields of the monopole theory)
are in the Higgs phase. We will show  that this leads to  formation of the
non-Abelian strings in the monopole theory whose tension is proportional to
a (fractional) power of $m_q$,
which in the original (quark) theory must be interpreted
as flux tubes of a ``chromoelectric" field.

To make our statement as clear as possible
it is worth comparing it with what was achieved in the Seiberg--Witten solution
\cite{SW1}. The low-energy limit of the ${\mathcal N}=2$ theory is a U(1) theory.
Upon dualization it becomes ${\mathcal N}=2$ SQED with
matter fields representing massless monopoles.  This theory is IR free.
Then the original theory is slightly deformed by a mass term of the adjoint chiral superfield. To the leading order this deformation does not
break ${\mathcal N}=2$ supersymmetry of low-energy SQED.
However, it changes the IR free regime into the Higss regime.
The monopoles condense, and BPS-saturated
light strings of the Abrikosov--Nielsen--Olesen type
\cite{ANO} ensue.  These are interpreted as the confining strings of the underlying electric theory.

Our goal is a similar scenario but in  non-Abelian version  in ${\mathcal N}=1$ theories.
The string we get is
non-Abelian and non-BPS. The worldsheet
theory (besides noninteracting translational and supertranslational moduli fields)
is nonsupersymmetric CP$(N-1)$ model, which has its own dynamics in the infrared.
In particular, kinks of the CP$(N-1)$ model are $N$-plets of a global symmetry.
In this sense our construction is closer to the QCD string, whatever it might be.

The paper is organized as follows.
In Sects.~\ref{quarkth} and~\ref{monopoleth} we introduce the
quark and the monopole theories related by extended Seiberg's duality,
and identify unbroken global symmetries. In Sect.~\ref{spectrum}
we study the pattern of physical scales relevant to both theories,
imposing the requirement of weak coupling in the monopole theory.
We show that this regime is self-consistent
and then study the spectrum of elementary excitations in the monopole
theory. In Sects.~\ref{strings} and~\ref{impli} we thoroughly discuss
the emergence of the non-Abelian strings and confined monopoles
in the monopole theory. The tension of the string is shown to be small
(vanishing in the limit $m_q\to 0$).
All hadrons related to these strings are light in the scale of $\Lambda_Q$.
Arguably, they represent a variety of light dual counterparts in the quark theory. The latter, being strongly coupled, tells us nothing about this rich set of low-lying states.
In Sect.~\ref{conf} which can be viewed as a conceptual culmination
 we continue the discussion of the light sectors
and suggest a quark-theory interpretation of the results
obtained in the monopole theory. Section \ref{conclu}
presents brief conclusions. Appendix A summarizes our notation.
In Appendix B we discuss peculiarities of the 't Hooft matching conditions
in our construction.

\section{Quark theory}
\label{quarkth}
\setcounter{equation}{0}

Our quark theory\,\footnote{A summary of our notation and conventions
is presented in Appendix A.} is \none SQCD with the gauge group
U$(N_c)={\rm SU}(N_c)\times {\rm U}(1)$ and
$N_c+N$ flavors of fundamental matter --- let us call it quarks.
As usual each quark flavor is described by two chiral superfields,
$Q$ and $\tilde Q$, one in the fundamental another in the antifundamental representation of U$(N_c)$.

We will endow
$N$ quark flavors (out of $N_c+N$) with equal mass terms $m_q$, while the remaining $N_c$  quark fields have vanishing mass terms. Our color/flavor notation
is as follows: the
quark supermultiplets are the
chiral superfields $Q^{kA}$ and $\tilde{Q}_{Ak}$ where $$k=1,...,N_c$$ and
$$A=1,..., (N_c+N)$$ are the color and flavor indices, respectively. The coupling constants
of SU$(N_c$) and U(1) gauge factors are denoted by $\left(g_{Q2}\right)^2$ and
$\left(g_{Q1}\right)^2$,
respectively. The subscript $Q$ will remind us that we deal with the quark theory.

In the supersymmetric vacuum
the scalar components of the quark multiplets $q$ and $\tilde{q}$ are subject to
$N^2_c$ real $D$-term conditions
\beq
q^{A}T^a \bar{q}_{A}-\bar{\tilde{q}}^{A}T^a\tilde{q}_{A}=0,\,\,\,\,(a=1,2,..., N_c^2-1)\,;\,\,\,\,\,\,
q^{A}T \bar{q}_{A}-\bar{\tilde{q}}^{A}T\tilde{q}_{A}=0\,,
\label{DtermQ}
\eeq
where  $T^a$ are generators of the SU($N_c$)
normalized by the condition
 $${\rm Tr}\,T^a T^b =\frac{1}{2}\,\delta^{ab}\,,$$
 while $T$ is the U(1) generator which we choose to be $T=1/2$.

The massless quark flavors
--- there are $N_c$ such flavors ---
develop VEVs breaking both SU$(N_c$) and U(1) gauge
groups. Then the theory is fully Higgsed. It has a vacuum manifold,  the Higgs
branch whose dimension is
\beq
{\rm dim} \, {\mathcal H}_Q=4N_c^2-N_c^2-N_c^2=2N_c^2 \,.
\label{dimHQ}
\eeq
Here we take into account the fact that we have $4N_c^2$ real variables $q^{kP}$ and
$\tilde{q}_{Pk}$ with the flavor index $P$,
$$P=1,...,N_c\,,$$
which describe massless squarks;
we subtract $N^2_c$ real $D$-term conditions and $N^2_c$ gauge phases
(for the U($N_c$) gauge group) eaten by the Higgs mechanism. In other words,
$2N_c^2-2$ real degrees of freedom (out of $4N_c^2$)
enter the  non-Abelian
gauge supermultiplets and acquire ``masses" $\sim \Lambda_Q$
where $\Lambda_Q$ is the dynamical scale parameter of the
quark theory. Two real degrees of freedom
enter the U(1) gauge supermultiplet and acquire masses
equal to that of the (Higgsed) photon.

The Higgs branch can be described in a gauge invariant way  by the meson
and baryon chiral moduli \cite{Seiberg1,Sdual,IS}\,\footnote{Note that
unlike \cite{Seiberg1} only the product
of the baryon operators $\tilde{B} B$
is gauge invariant in the case of the U($N_c$) gauge group.}
\beq
\langle \tilde{q}_S q^P\rangle , \qquad \tilde{B} B \,,
\label{ginv}
\eeq
subject to the condition
\beq
{\rm det}\, \langle \tilde{q}_S q^P\rangle -\tilde{B} B
=\Lambda_Q^{2N_c-N}m_q^N \,,
\label{condition}
\eeq
where
\beq
B= \frac1{N_c !}\,\varepsilon_{k_1 ... k_{N_c}}  \, q^{k_1 1}...q^{k_{N_c} N_c},
\qquad \tilde{B}= \frac1{N_c !}\,
\varepsilon^{k^1 ... k^{N_c}} \, \tilde{q}_{1k_1 }...\tilde{q}_{N_c k_{N_c} }\,.
\label{baryonops}
\eeq
One can view
\beq
\Lambda_{Q, le} =
\left(\Lambda_Q^{2N_c-N}m_q^N \right)^{1/(2N_c)}
\eeq
as a scale parameter of the effective low-energy theory emerging at momenta
below $m_q$. However, this parameter will play no role in what follows.

Using ${\rm SU}(N_c)\times {\rm SU}(N_c)$ flavor rotations
we can always transform the matrix of the vacuum expectation values
$\langle \tilde{q}_S q^P\rangle$ to a diagonal form,
\beq
\langle \tilde{q}_S q^P\rangle \to\delta_S^P\, {\mathcal Q}_P\,,
\label{higgsi}
\eeq
where $N_c$ parameters $ {\mathcal Q}_P$
determine the position of the vacuum on the vacuum manifold.
Generically  ${\mathcal Q}_1\neq {\mathcal Q}_2\neq ... \neq {\mathcal Q}_{N_c}$.
We will assume that all ${\mathcal Q}$'s are nondegenerate
and of the same order of magnitude,
\beq
{\mathcal Q}_1\sim  {\mathcal Q}_2\sim ... \sim {\mathcal Q}_{N_c} \sim {\mathcal Q}\,.
\label{higgsip}
\eeq
Finally, we will assume that
\beq
{\mathcal Q} \ll \Lambda_Q^2\,,
\label{vevsm}
\eeq
see below for a more detailed discussion. Equation (\ref{vevsm})
implies the strong coupling regime.

\vspace{4mm}

Now, let us make a step back and set $m_q=0$. Then the quark theory
at the Lagrangian level  has
\beq
{\rm SU}(N_c+N)_L\times {\rm SU} (N_c+N)_R\times {\rm U}(1)_R\times
{\rm U}(1)_A
\label{glgrouptree}
\eeq
global symmetry. The  axial U(1)$_A$ symmetry  is anomalous.
The  U(1)$_R$
symmetry is chosen to be non-anomalous with respect to non-Abelian
gauge bosons. However, it appears to be anomalous with respect
to the U(1) gauge bosons, see Appendix B.
Note the absence of the global baryon U(1) symmetry. The baryon charge
is gauged in our theory.

If $m_q=0$ this global group is broken by quarks VEVs down to
\beq
{\rm SU}(N)_L\times {\rm SU}(N)_R  \,,
\label{glgroupmzer}
\eeq
The number of broken generators is  $2N_c^2+4NN_c$.

The global symmetry (\ref{glgroupmzer}) is broken by the quark mass
$m_q\neq 0$
down to a diagonal
\beq
{\rm SU}(N)\,.
\label{global}
\eeq
When a small mass $m_q$ is switched on, some of the
spontaneously broken
generators are broken explicitly too and, thus, interpolate
pseudo-Goldstone (pG) rather than the Goldstone states. The number of
true massless states is given by
the dimension of the Higgs branch (\ref{dimHQ}) while the number of the
pseudo-Goldstone  states is
\beq
N_{\rm pG}=4NN_c.
\label{numbpG}
\eeq
These pG states (interpolated
by ${\tilde Q}_S\,Q^{\dot K}$ where ${\dot K} = N_c+1,..., N_c+N$
while $S=1,..., N_c$, see Appendix A) have masses
\beq
m_{\rm pG}\sim m_q.
\label{masspG}
\eeq
The pseudo-Goldstones
${\tilde Q}_S\,Q^{\dot K}$ are in the fundamental representation
of the global unbroken ${\rm SU}(N)$.

Other light states
 are not easily seen on this side of duality
but can be inferred from its dual description (see Sect.~\ref{spectrum}),
in particular, a vector multiplet and ``pions" $\tilde{q}_{\dot L} q^{\dot K}$
in the adjoint representation of the global  ${\rm SU}(N)$.

\vspace{4mm}

The first coefficient of the $\beta$ function of the quark theory is
\beq
b_Q=2N_c-N\,.
\label{bQ}
\eeq
To make it positive (so the theory is asymptotically free) it is sufficient to
have $N<2N_c$. However, we will limit ourselves to even smaller values
of $N$.
We will consider the quark theory well below the left edge of the conformal window \cite{Shifman1Rev},
\beq
N< \frac{N_c}{2} \,.
\label{belowwind}
\eeq
Moreover, we will assume the quark  mass terms to be the smallest
dimensional parameter in the quark theory,
\beq
m_q\ll  \sqrt{\mathcal Q} \,\ll \Lambda_Q\,.
\label{smallmq}
\eeq
This condition means that $N$ quark flavors with vanishing VEVs are dynamical and do
not decouple, while the condition (\ref{vevsm}) ensures that the quark theory is in the
strong coupling regime. If the quark VEVs were much larger than $\Lambda_Q^2$
the  quark theory would be in the Higgs phase at weak coupling. Instead, at
small quark VEVs the theory is in the strong coupling regime and,
although the Higgs and confinement phases are analytically connected in theories with
fundamental matter \cite{FrSh},
it is more convenient to speak of the quark theory in terms of
confinement.

\section{Monopole  theory}
\label{monopoleth}
\setcounter{equation}{0}

To begin with, let us put $m_q=\mathcal Q =0$ in the quark theory.
In this case we  can just
follow Seiberg \cite{Sdual} and accept that the dual description of the original
quark theory (Seiberg's ``magnetic  dual") is given by an \none supersymmetric gauge theory with certain matter fields. The gauge group  is U$(N)$.
The \none vector multiplets consist of the  U(1)
gauge fields $A_{\mu}$ and SU($N$)  gauge field $A^a_{\mu}$,
(here $a=1,..., N^2-1$) and  their Weyl fermion superpartners
$\lambda_{\alpha}$, and
$\lambda^{a}_{\alpha}$.  The spinorial index of $\lambda$'s runs over $\alpha=1,2$.

There are $N+N_c$ flavors in the monopole theory. From the standpoint
of the  monopole theory {\em per se} each flavor is represented by
a pair of  superfields $h^{kA}$ and $\tilde{h}_{Ak}$ with the lowest (scalar)
components
$h^{kA}$ and $\tilde{h}_{Ak}$  and
the  Weyl fermions $\psi^{kA}$ and
$\tilde{\psi}_{Ak}$,
 all in the fundamental representation of the  SU($N$)  gauge group.
Here $k=1,..., N\,\,$ is the color index
while $A$ is the flavor index, $A=1,...,(N+N_c)$. Seiberg termed
these $h$ fields ``dual quarks."
From the standpoint of the original quark theory the
dual quarks are to be interpreted as monopole fields.

In addition, the monopole theory has a
gauge neutral ``mesonic" field $M^B_A$
which is locally related to the $ \tilde{Q}_A Q^B$ composite chiral operator of the original quark theory
\cite{Sdual,IS},
\beq
\tilde{Q}_A Q^B=\kappa \,M_A^B\,,
\label{QQM}
\eeq
where $\kappa$ is an energy scale to be determined below.

The mesonic field is coupled to the monopole fields via the superpotential \cite{Sdual,IS}
\beq
W_{\rm Yukawa}= \tilde{h}_{Ak}\,h^{kB}\,M^A_B \,.
\label{Yuksup}
\eeq

The mass term $m_q\,\tilde{Q}_{\dot K} Q^{\dot K} $ of the quark theory takes the form
of a linear in $M$ superpotential
\beq
 W_{\rm linear}=-\,\frac{1}{2}\,\xi\, M_{\,\,\dot K}^{\dot K} \,.
\label{linear}
\eeq
From Eq.~(\ref{QQM}) we find the following relation between the parameter $\xi$ in the linear
superpotential (\ref{linear}) and the quark masses in the original theory:
\beq
\xi=-2\kappa \, m_q\, .
\label{xim}
\eeq
The bosonic part of the action of the monopole theory (in the limit
$m_q=\mathcal Q =0$) is
\beqn
S&=&\int d^4x \left[\frac1{4\left( g_{M2}\right)^2}
\left(F^{a}_{\mu\nu}\right)^2 +
\frac1{4\left( g_{M1}\right)^2}\left(F_{\mu\nu}\right)^2+
{\rm Tr}\,\left|\nabla_{\mu}
h\right|^2 + {\rm Tr}\,\left|\nabla_{\mu} \bar{\tilde{h}}\right|^2
\right.
\nonumber\\[4mm]
&+& \frac2{\gamma} {\rm Tr}\,\left|\pt_{\mu} M\right|^2
+\frac{g^2_{M2}}{2}
\left(
 {\rm Tr}\,\bar{h}\,T^a h -
{\rm Tr}\,\tilde{h} T^a\,\bar{\tilde{h}}\right)^2
\nonumber\\[3mm]
&+& \frac{g^2_{M1}}{8}
\left({\rm Tr}\,\bar{h} h - {\rm Tr}\,\tilde{h} \bar{\tilde{h}}
\right)^2+
 {\rm Tr}|hM|^2 +{\rm Tr}|\bar{\tilde{h}}M|^2
\nonumber\\[3mm]
&+&
\left.
\frac{\gamma}{2}\left|\tilde{h}_A h^B -\frac{1}{2}\delta_A^{\dot K} \delta_{\dot K}^B\,\xi\right|^2
\right\}
\,,
\label{mmodel}
\eeqn
where the covariant derivatives are defined as\,\footnote{For further explanations on our notation see Appendix A.}
\beq
\nabla_\mu=\partial_\mu -\frac{i}{2}\; A_{\mu}
-i A^{a}_{\mu}\,T^a\,.
\label{defnabla}
\eeq
The trace in (\ref{mmodel}) runs over all flavor indices. In addition
to the dual gauge couplings
$g_{M1}$ and $g_{M2}$ for the U(1) and SU$(N)$ factors, respectively, we introduced
the coupling constant  $\gamma$ for the $M$ field.

Let us note that the duality pairs were found in
Ref. \cite{Sdual}  for theories with the gauge groups SU$(N)$ rather than U$(N)$.
To generalize Seiberg's duality to the latter case we gauge the U(1) global baryon symmetry present on both sides of Seiberg's duality.

If in the original quark theory the
quark fields have the baryon charge 1/2, then the baryon charges of
the monopole fields in the
monopole theory are ${N_c}/{(2N)}$ (see Refs.~\cite{Sdual,IS}). If we still want to keep
the corresponding couplings identical we must choose
\beq
g_{M1}=\frac{N_c}{N}\,g_{Q1}\,.
\label{g1}
\eeq
Then on both sides of Seiberg's duality the U(1) charge is 1/2 being measured
in the units of the appropriate gauge coupling. Then the definition
 (\ref{defnabla}) stays intact.

The SU$(N)$ gauge coupling constant $g_{M2}$ in (\ref{mmodel}) is determined by the scale $\Lambda_M$
of the monopole theory. The latter
is related to the scale $\Lambda_Q$ of the quark theory as
\cite{Sdual,IS}
\beq
\Lambda_Q^{2N_c-N}\Lambda_M^{2N-N_c}=(-1)^{N}\kappa^{N_c+N}\,.
\label{lrelation}
\eeq
As we will see shortly, $\Lambda_M$ is the largest parameter
in our analysis, $\Lambda_M\gg\Lambda_Q$, but it will play no role
since the monopole theory in the limit $m_q=\mathcal Q =0$ (and with
$N$ satisfying the condition (\ref{belowwind}))  is infrared rather than
asymptotically free; it lies to the right of the right edge of the conformal window.
What will be important for dynamical considerations is an effective low-energy
parameter  $\Lambda_{M,le}$ which will emerge after
$m_q\neq 0$ and $\mathcal Q \neq 0$ are taken into account.

It is natural to assume that
\beq
\gamma \sim 1\,.
\eeq

\vspace{4mm}

Now let us switch on $m_q\neq 0,\,\,\,\mathcal Q \neq 0$
in the quark theory
and  discuss the vacuum structure of the monopole theory.
The linear in $M$ superpotential which reflects $m_q\neq 0$
triggers spontaneous breaking
of the U$(N)$ gauge symmetry. The vacuum expectation values
of the $h$  fields can be chosen as
\beq
\langle h^{k\dot{K}}\rangle = \langle \bar{\tilde{h}}^{k\dot{K}}\rangle=
\sqrt{\frac{\xi}{2}}\, \left(
\begin{array}{ccc}
1 & 0 & ...\\
... & ... & ... \\
... & 0 & 1  \\
\end{array}
\right),\qquad
\langle h^{kP}\rangle = \langle \bar{\tilde{h}}^{kP}\rangle =0\,,
\label{qvev}
\eeq
{\em up to gauge rotations}.

Furthermore, Higgsing the quark theory (\ref{higgsi}), (\ref{higgsip})
manifests itself in the monopo\-le theory as
nonvanishing
VEVs of the $M^P_S$
fields  related to VEVs
of $ \tilde{Q}_S Q^P$ via (\ref{QQM}),
\beq
\langle M^P_S\rangle =
\left(
\begin{array}{ccc}
{\mathcal M}_1 & 0 & ...\\
... & ... & ... \\
... & 0 & {\mathcal M}_{N_c}  \\
\end{array}
\right)
\label{Mvev}
\eeq
where
${\mathcal M}_1\sim {\mathcal M}_2\sim ...\sim {\mathcal M}_{N_c}\sim \mathcal M$
and we introduced a common scale $\mathcal M$.
The above nonvanishing VEVs
make the first $N_c$ monopole flavors massive by virtue
of $W_{\rm Yukawa}$. If we descend below $\mathcal M$
the massive flavors can be integrated out. What remains is a U$(N)$
gauge theory with $N$ flavors which is asymptotically free.
The scale of this  theory $\Lambda_{M,le}$ is defined via the relation
\beq
\Lambda^{2N}_{M,le}=\Lambda_M^{2N-N_c}\mathcal{M}^{N_c}\,,
\label{Lambda}
\eeq
which, in turn, can be expressed in terms of the
quark theory scale,
\beq
\Lambda_Q^{2N_c-N}\Lambda^{2N}_{M,le} =(-1)^{N}\mathcal{M}^{2N_c+N}
\label{llQrelation}
\eeq
by invoking Eq.~ (\ref{lrelation}). As we will see shortly, the scale
$\Lambda_{M,le} $ lies between $m_q$ and $\sqrt\xi$.
This guarantees that the monopole theory is weakly coupled.

\vspace{4mm}

Other $M$ fields do not condense,
\beq
\langle M^{\dot K}_{\dot L}\rangle =\langle M^P_{\dot L}\rangle
=\langle M^{\dot K}_S\rangle=0 \,.
\label{veva}
\eeq

\vspace{4mm}

The color-flavor locked form of the quark VEVs in
Eq.~(\ref{qvev}) and the vanishing of VEVs
 in (\ref{veva}) result in the fact that, while the theory is fully Higgsed, a diagonal
SU$(N)_{C+F}$ symmetry survives as a global symmetry. Namely, the global rotation
\beq
h\to UhU^{-1},\qquad \bar{\tilde{h}}\to U\bar{\tilde{h}}U^{-1}, \qquad M\to U^{-1}MU
\label{c+f}
\eeq
is not broken by the VEVs (\ref{qvev}) and (\ref{Mvev}). Here $U$ is an arbitrary
 matrix from SU($N$).
We write the dual quark (monopole) fields $h^{k\dot{K}}$,
$\tilde{h}_{\dot{K}k}$ and $M^{\dot K}_{\dot L}$ with indices $ k
=1,..., N$ and  ${\dot K},\,\,{\dot L}= N_c+1,..., N_c+N$ as $N\times N$
matrices; the matrices $U$ act on these indices.  This is a particular
case  of the Bardak\c{c}\i--Halpern mechanism \cite{BarH}.

Classically,
in addition to the unbroken global SU$(N)_{C+F}$ symmetry, there is also
a chiral ${\rm U}(1)_{R''}$ symmetry which survives the breaking induced by
VEVs (\ref{qvev}) and (\ref{Mvev}), see Appendix B.
However, it turns out to be anomalous with respect to the U(1) gauge fields.
Therefore, the global unbroken symmetry of the monopole theory
\beq
{\rm SU}(N)_{C+F}
\label{globalM}
\eeq
is the same as the symmetry of the quark theory (\ref{global}).
In Appendix B we check duality  by demonstrating that the 't Hooft
anomaly matching conditions  in the  quark and monopole
theories are satisfied. There are certain peculiarities
since the matching looks different at ``high energies"
(i.e. when the momentum $q$ flowing in the axial curent
satisfies $q^2\gg \xi$) and ``low energies" ($q^2\ll \xi$).
We check both limits.

The SU$(N)_{C+F}$  global symmetry of the theory is spontaneously
broken on strings, which gives rise to the
orientational zero modes \cite{ABEKY} of the $Z_N$ strings in the model (\ref{mmodel}).

Below we  assume that
the original quark theory has the same
low-energy  physics as the monopole theory (\ref{mmodel}). By low energies we mean
scales  $\sqrt\xi$ and below.

\section{Elementary excitations in  the monopole  theory}
\label{spectrum}
\setcounter{equation}{0}

First we observe that the fields $M_S^P$ can develop VEVs; thus, the dimension
of the Higgs branch in the monopole theory
\beq
{\rm dim}\, ({\mathcal H}_M)=2N_c^2
\label{dimHM}
\eeq
agrees with the one in the quark theory (\ref{dimHQ}).
As a result, $2N_c^2$ (real) fields $M_S^P$ are
massless.

Now we will demonstrate that the scale of VEVs of the $M_S^P$ fields is the largest
relevant parameter in
the monopole theory. In particular, it is much larger than the effective scale
$\Lambda_{M,le}$ of the monopole theory.
Equation (\ref{higgsip}) implies that all parameters ${\mathcal M}_S$
in Eq.~(\ref{Mvev})
are of the same order,
\beq
{\mathcal M}_1\sim {\mathcal M}_2\sim ...\sim {\mathcal M}_{N_c}\sim \mathcal M \,.
\eeq
Due to the Yukawa interactions (\ref{Yuksup}) the
flavors $h^{kP}$ ($\tilde{h}_{Pk}$) become massive, with masses
$m(h^P)\sim \mathcal M$,
and decouple below this scale. Integrating them out in the superpotential (\ref{Yuksup}) produces an effective
low-energy  superpotential
\beq
W_{M,le}= \tilde{h}_{{\dot K}k}\,h^{k{\dot L}}\,\left[M^{\dot K}_{\dot L}-
\frac{M^{\dot K}_P M^P_{\dot L}}{\mathcal M}\right] \,.
\label{LEsup}
\eeq
This superpotential gives small masses to $4NN_c$ (real) off-diagonal fields
$M^{\dot K}_P$,
$M^P_{\dot K}$. To see that this is indeed
the case please observe that the monopole fields $h^{k{\dot K}}$
develop VEVs $\sim \sqrt{\xi}$
(see Eq.~(\ref{qvev})). Then the second term in (\ref{LEsup}) implies
\beq
m(M^P_{\dot K})\sim m(M^{\dot K}_P)\sim \frac{\xi}{\mathcal M }\,.
\label{light}
\eeq
We see that the number of the off-diagonal  $M$ fields that are light coincides with the number
of the pseudo-Goldstone fields $\tilde{q}_{\dot L} q^P$ in the quark theory, see (\ref{numbpG}). Requiring
their masses to be the same we get
\beq
\kappa = \mathcal M \,,
\label{kappa}
\eeq
where we used (\ref{masspG}), (\ref{light}) and (\ref{xim}). This fixes the
so-far  unknown
coefficient $\kappa$. All three scales of the monopole theory, namely $\Lambda_M$,
$\xi$ and $\mathcal M$ are now fixed in terms of three scales of the quark theory
$\Lambda_Q$, $m_q$ and
\beq
\langle \tilde{Q}Q\rangle \sim \mathcal Q \sim {\mathcal M}^2\,.
\eeq

As was mentioned,
below the scale $\mathcal M$  the
effective-low energy theory is the U$(N)$ gauge theory with $N$ flavors
of the $h^{k\dot{K}}$ fields  supplemented by the mesonic
(gauge-singlet) field $M$, out of which $M_S^P$ are massless,
$M^{\dot K}_P$ and $M^P_{\dot K}$ have masses $\sim m_q$ and
$M^{\dot K}_{\dot L}$ have masses $\sim \sqrt\xi$, see below.  We consider
this theory
in the weak coupling regime imposing the condition
\beq
\xi\gg
\Lambda_{M,le}^2\,.
\label{Lxi}
\eeq
In terms of the quark theory
scales this condition can be rewritten by virtue of
Eq.~(\ref{llQrelation}) as
\beq
{\mathcal M}^{2N_c}\ll
\Lambda_Q^{2N_c-N}m_q^N.
\label{weakcoupM}
\eeq
We see that in order
to keep the monopole theory at weak coupling, the  scale $\mathcal M $
cannot be  too large.

To elucidate the meaning of this condition,  following \cite{IS}  we
relate the baryon operators in the  quark theory $B$ and $\tilde{B}$ to
the baryon operators
in the monopole theory
\beqn
B
&=&
 q^1...q^{N_c}= \left[-(-1)^N \kappa^{-N}\Lambda_Q^{2N_c-N}\right]^{\frac12}\,h^1...h^N \, ,
\nonumber\\[3mm]
\tilde{B}
&=&
 \tilde{q}_1...\tilde{q}_{N_c}=
\left[-(-1)^N \kappa^{-N}\Lambda_Q^{2N_c-N}\right]^{\frac12}
\,\tilde{h}^1...\tilde{h}^N\,.
\label{Bb}
\eeqn
The right-hand sides of these expressions
 have nonvanishing VEVs in the vacuum of the monopole theory,
see (\ref{qvev}). Note, however,  that the $h$ field VEVs
are given by (\ref{qvev}) only classically. We expect corrections
in $\langle h ... h\rangle$ of order of $\Lambda_{M,le}$ to
the product of the classical expectation values (\ref{qvev}).
These corrections are
 small provided (\ref{Lxi}) is satisfied.

Substituting  VEVs (\ref{qvev}) in (\ref{Bb})
and ignoring the above corrections we get for  the baryon operator VEV
 in the quark theory
\beq
\langle \tilde{B}B\rangle
=-\,\Lambda_Q^{2N_c-N}m_q^N\, ,
\label{barQ}
\eeq
where we
used Eq.~(\ref{xim}). We see that the relation (\ref{condition}) of the
quark theory is saturated to the leading order by the baryon operator.
Quantum corrections to VEVs (\ref{qvev})  substituted into
(\ref{condition}) allow for
\beq
{\rm det}\, \langle \tilde{Q}_P
Q^S\rangle \sim \mathcal{M}^{2N_c} \ll \Lambda_Q^{2N_c-N}m_q^N =
-\,\langle \tilde{B}B\rangle.
\label{barmeson}
\eeq
Thus, the weak coupling
regime in the monopole theory corresponds to the baryon dominated vacuum in
the quark theory.

\begin{figure}[h]
\epsfxsize=14cm
\hspace{4cm}\epsfbox{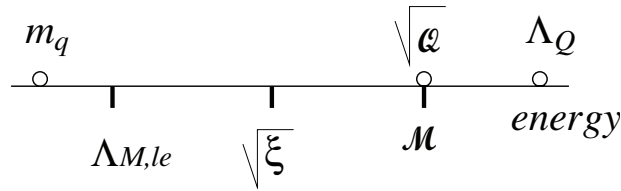}
%\centerline{\epsfbox{mscale.eps}}
\caption{\footnotesize
Scales of the quark (open points)  and monopole (dashes) theories.}
\label{figscales}
\end{figure}

Relevant scales of the quark and monopole theories are shown
in Fig.~\ref{figscales}.

Now let us prove the statement made above regarding the states with masses of the
order of $\sqrt{\xi} $.
Since
both the U(1) and SU($N$) gauge groups are broken by the $h^{\dot K}$
field condensation, see Eq.~(\ref{qvev}),
all gauge bosons become massive. From (\ref{mmodel}) we get for the U(1)
gauge boson mass (dual ``photon")
\beq
m_{\rm ph}=g_{M1}\sqrt{\frac{N}{2}} \,\sqrt{\xi}\,.
\label{phmass}
\eeq
The $(N^2-1)$ gauge bosons of the SU($N$) group
(dual ``W bosons")
acquire a common mass
\beq
m_{W}=g_{M2}\,\sqrt{\xi}\,.
\label{wmass}
\eeq
This is typical of the Bardak\c{c}\i--Halpern mechanism.
Needless to say, $N^2$ vector states with masses $\sim\sqrt\xi$
must appear in the quark theory too.

To get the scalar boson masses we expand the potential in (\ref{mmodel})
near the vacuum (\ref{qvev}),  (\ref{Mvev}) and diagonalize the
corresponding mass matrix. The $N^2$ components of the
$2\,N^2$-component\,\footnote{We mean here  {\em real} components.}
scalar $(h-\bar{\tilde{h}})^{k\dot{K}}/\sqrt{2}$
are eaten by the Higgs mechanism for the U(1) and SU($N$)
gauge groups. Another $N^2$ components are split as follows:
one component acquires the mass (\ref{phmass}). It becomes
 a scalar component of  a massive \none vector U(1) gauge multiplet.
The remaining $N^2-1$ components acquire masses (\ref{wmass}) and become
scalar superpartners of the SU($N$) gauge bosons in the \none massive gauge
supermultiplet.

The fields $M^{\dot K}_{\dot L}$ and $(h+\bar{\tilde{h}})^{k{\dot K}}/\sqrt{2}$ form chiral multiplets.
Namely, the states proportional to the unit $N\times N$ matrix
(associated with U(1)) acquire masses
\beqn
m_{{\rm U}(1)}^{M} &=&\sqrt{\frac{\gamma N\xi}{4}}\,,
\nonumber\\[3mm]
m_{{\rm U}(1)}^{h} &=&  \sqrt{\frac{\gamma N\xi}{4}}\,,
\label{U1mass}
\eeqn
respectively, while  the traceless parts
of $M^{\dot K}_{\dot L}$ and $(h+\bar{\tilde{h}})^{k{\dot K}}/\sqrt{2}$
(associated with the SU($N$) sector) have masses
 \beqn
m_{{\rm SU}(N)}^{M} &=& \sqrt{\frac{\gamma\xi}{2}}\,,
\nonumber\\[3mm]
m_{{\rm SU}(N)}^{h} &=&  \sqrt{\frac{\gamma\xi}{2}}\,.
\label{SUNmass}
\eeqn

Other states with masses of the order of $\mathcal M $ are much heavier and we do not
include them in the low-energy spectrum.  To reiterate,
in the monopole theory the low-energy spectrum includes the Goldstone and
pseudo-Goldstone states
(with masses (\ref{light})), as well as the states with masses
 $\sim \sqrt{\xi}$, ``elementary" states discussed above and composite states to be discussed
below.  In Sect.~\ref{strings} we will discuss formation
of non-Abelian strings in the monopole theory. These strings produce (extra)
non-perturbative
states in the monopole theory with masses  $\sim\sqrt{\xi}$.
Needless to say, all these states from the low-energy sector of the monopole theory
are presumed to exactly match the low-energy spectrum of the quark theory.
This is the statement of extended duality.

\section{Non-Abelian strings}
\label{strings}
\setcounter{equation}{0}

Non-Abelian strings were shown to emerge at weak coupling
in \ntwo  supersymmetric gauge theories \cite{HT1,ABEKY,SYmon,HT2}.
Recently they were also found \cite{Mmodel} in
\none  super\-symmetric theory with the U$(N)$ gauge group and $N$
fundamental matter multiplets supplemented by a ``mesonic'' field
$M^{\dot K}_{\dot L}$
\,(${\dot K},\,\, {\dot L}=N_c+1,..., N_c+N$).

The $M$ model of Ref.~\cite{Mmodel} is
a close relative of the monopole theory we consider here.
The two models differ in that there are extra components of the
$M$ fields in the monopole theory, namely the Goldstones $M^P_S$ and
pseudo-Goldstones $M^{\dot K}_P$, $M^P_{\dot L}$.
The impact of their presence is rather unimportant, as we will see below.
Extra monopole
fields $h^{kP}$ present in the monopole theory are heavy and  can be
ignored.

\subsection{Non-BPS strings}

Since the $M$ field does not enter the classical string
solution, the explicit solution for the non-Abelian strings found
in the  $M$ model  \cite{Mmodel} can be readily adjusted to fit
the monopole theory at hand.
The light components of the $M$ field which have no counterparts
in the $M$ model will show up only at the quantum level.

A consequential distinction of the monopole theory   from the $M$ model
is that the $\xi$ parameter which triggers the $h$-field condensation is introduced via
superpotential (\ref{linear})
rather than through the FI $D$ term as in the original $M$ model.
Introducing $\xi$ through the superpotential is a viable alternative.
In this case we will speak of
$M_F$ model while the original $M$ model can be termed
$M_D$.

Note that the $M_F$ model strings
are necessarily non-BPS \cite{GoS},
while in the $M_D$ model we deal with the  BPS-saturated strings.
This means that the worldsheet  theory on the strings
will  be non-supersymmetric in the case at hand.

The bosonic part of the action of the $M_F$ model  is given by
(\ref{mmodel}), with the flavor indices running only over $N$ values
$A,B\to {\dot K},{\dot L}=N_c+1,..., N_c+N$. This is shown in more
detail in Appendix A.

The scalar fields   involved in the string solution are
\beq
h^{k{\dot K}}=\bar{\tilde{h}}^{k{\dot K}}\equiv \frac1{\sqrt{2}}\varphi^{k{\dot K}}
\, .
\label{hth}
\eeq
With this substitution the {\em ansatz}  for the solution for the
elementary non-Abelian
strings (in the singular gauge) becomes \cite{Mmodel}
\beqn
\varphi &=& \frac1N[(N-1)\phi_2 +\phi_1] +(\phi_1-\phi_2)\left(
n\,\cdot n^*-\frac1N\right) ,
\nonumber\\[3mm]
A^{{\rm SU}(N)}_i &=& \left( n\,\cdot n^*-\frac{1}{N}\right)
\varepsilon_{ij}\, \frac{x_i}{r^2}
\,
f_{NA}(r) \,,
\nonumber\\[3mm]
A^{{\rm U}(1)}_i &=& \frac1N
\varepsilon_{ij}\, \frac{x_i}{r^2} \, f(r) \, ,
\nonumber\\[3mm]
    M^{\dot K}_{\dot L} & = & M^{\dot K}_P= M^P_{\dot L}=0\,,
\label{str}
\eeqn
where $r$ is the distance from the string axis to the given point in the
orthogonal plane. Moreover,
$n^l$ is a set of complex scalar fields forming the fundamental representation of
SU($N$) subject to the constraint
\beq
 n^*_l n^l =1\,,
\label{unitvec}
\eeq
($l=1, ..., N$ is the SU$(N)$ index). In Eq.~(\ref{str}) for brevity we suppress
all SU$(N)$  indices. Varying $n^l$ we change the orientation of the string
flux  in the
non-Abelian color subgroup SU$(N)$. It is associated with the orientational zero modes of the non-Abelian string.

The profile functions for the scalar fields $\phi_1(r)$, $\phi_2(r)$ and for the gauge fields
$f(r)$, $f_{NA}(r)$ satisfy the following boundary conditions:
\beqn
&& \phi_{1}(0)=0,
\nonumber\\[2mm]
&& f_{NA}(0)=1,\;\;\;f(0)=1\,,
\label{bc0}
\eeqn
at $r=0$, and
\beqn
&& \phi_{1}(\infty)=\sqrt{\xi},\;\;\;\phi_2(\infty)=\sqrt{\xi}\,,
\nonumber\\[2mm]
&& f_{NA}(\infty)=0,\;\;\;\; \; f(\infty) = 0
\label{bcinfty}
\eeqn
at $r=\infty$.

To see that the strings in the $M_F$ model  are not BPS
it is sufficient to note that the masses of the scalar field $\varphi$ given in
the second lines in Eqs.~(\ref{U1mass}) and (\ref{SUNmass})
are  not the same as the gauge boson masses (\ref{phmass}) and (\ref{wmass})
(for generic values of the coupling constant $\gamma$).
In the $M_D$ model the scalars involved in the string solution are in fact the
scalar superpartners of the gauge bosons from \none massive vector supermultiplets.
The equality of masses of the scalar and gauge fields is ensured and protected by
supersymmetry. This is the reason why the $M_D$ model-strings are
BPS-saturated \cite{Mmodel}.

If we considered a special set
of the coupling constants,
\beq
g_{M1}^2=g_{M2}^2=\gamma/2\equiv g^2\,,
\label{gammag}
\eeq
the equality of masses of the scalar
fields (\ref{U1mass}) and (\ref{SUNmass}) and the gauge fields (\ref{phmass}) and
(\ref{wmass}) would be guaranteed at the classical level. In this case the  string  profile functions
would  satisfy the following first-order equations \cite{ABEKY,Mmodel}:
\beqn
&&
r\frac{d}{{d}r}\,\phi_1 (r)- \frac1N\left( f(r)
+ (N-1) f_{NA}(r) \right)\phi_1 (r) = 0\, ,
\nonumber\\[4mm]
&&
r\frac{d}{{ d}r}\,\phi_2 (r)- \frac1N\left(f(r)
-  f_{NA}(r)\right)\phi_2 (r) = 0\, ,
\nonumber\\[4mm]
&&
-\frac1r\,\frac{ d}{{ d}r} f(r)+\frac{g^2 N}{4}\,
\left[\left(\phi_1(r)\right)^2 +(N-1)\left(\phi_2(r)\right)^2-N\xi\right] =
0\, ,
\nonumber\\[4mm]
&&
-\frac1r\,\frac{d}{{ d}r} f_{NA}(r)+\frac{g^2}{2}\,
\left[\left(\phi_1(r)\right)^2 -\left(\phi_2(r)\right)^2\right]  = 0
\, .
\label{foe}
\eeqn
For generic values of the coupling constants the string profile functions
satisfy the second-order equations. The condition
(\ref{gammag}), even being imposed at the classical level, will be certainly destroyed by loop corrections.

Assuming (\ref{gammag}) we would find
the tension of the elementary string
\beq
T_{\rm string} =2\pi\,\xi + {\rm quantum\,\,\, corrections}\,.
\label{ten}
\eeq
The $2\pi\,\xi$ is no longer exact. Without (\ref{gammag}), and with no
explicit solution of the second-order equations
we can only say that the string tension $T_{\rm string}\propto \xi $.
The non-Abelian string discussed above 
presents an SU$(N)$ rotation of the $Z_N$ strings \cite{ABEKY}.
The rotation parameters are determined by $n_l$, ($l=1,2,...N$), see Eq.~(\ref{str}).
The $Z_N$ string carries a combination of magnetic fluxes:
the magnetic flux of the U(1) field as well as that of the non-Abelian fields.
At the classical level the color orientation of the
non-Abelian fluxes is fixed in the Cartan subgroup,  in a well defined manner. 
At the quantum level the color orientation of the
non-Abelian fluxes strongly fluctuates, in accordance with dynamics
of the worldsheet CP$(N-1)$ model, so that the entire SU$(N)$
group space is spanned. 

To conclude this section let us note a somewhat related development:
{\em non}-BPS non-Abelian strings were recently
considered in metastable vacua of a dual description of \none SQCD at $N_f>N$ in
Ref.~\cite{Jmeta}.

\subsection{Worldsheet theory}

To derive a  worldsheet  theory describing orientational moduli
 $n^l$ of the non-Abelian string we follow
Refs.~\cite{ABEKY,SYmon,GSY05}, see also the  review paper \cite{SYrev}.

Assume  that the orientational collective coordinates $n^l$
are slowly varying functions of the string worldsheet coordinates
$x_k$, $k=0,3$. Then the moduli $n^l$ become fields of a
(1+1)-dimensional sigma model on the worldsheet. Since
the vector $n^l$ parametrizes the string zero modes,
there is no potential term in this sigma model.

To obtain the   kinetic term  we substitute our solution, which depends
on the moduli $ n^l$, in the action (\ref{mmodel}), assuming  that
the fields acquire a dependence on the coordinates $x_k$ via $n^l(x_k)$.
Then we arrive at the CP$(N-1)$  sigma model (for details see  the review
paper \cite{SYrev}),
\beq
S^{(1+1)}_{{\rm CP}(N-1)}= 2 \beta\,   \int d t\, dz \,  \left\{(\pt_{k}\, n^*
\pt_{k}\, n) + (n^*\pt_{k}\, n)^2\right\}\,,
\label{cp}
\eeq
where the coupling constant $\beta$ is given by a normalizing integral $I$
defined in terms of the string profile functions,
\beq
\beta= \frac{2\pi}{g_{M2}^2}\,I.
\label{betag}
\eeq
The two-dimensional coupling constant is determined by the
four-dimensional non-Abelian coupling.

Using the first-order equations for the string profile functions (\ref{foe})
one can see that
the integral $I$  reduces to a total derivative and is given
by the string  flux  determined by $f_{NA}(0)=1$, namely $I=1$. However, for the
non-BPS string in the problem at hand we certainly expect corrections to this
classical BPS result. In particular, we expect that generically $I$ acquires a dependence
on $N$ and coupling constants.

The relation between the four-dimensional and two-dimensional coupling
constants (\ref{betag}) is obtained  at the classical level. In quantum theory
both couplings run. So we have to specify a scale at which the relation
(\ref{betag}) takes place. The two-dimensional CP$(N-1)$ model is
an effective low-energy theory appropriate for describing
string dynamics  at low energies,  much lower than the
inverse string thickness which, in turn, is given by $g_{M2}\sqrt{\xi}$. Thus,
$g_{M2}\sqrt{\xi}$ plays the role of a physical ultraviolet (UV) cutoff in
(\ref{cp}).
This is the scale at which Eq.~(\ref{betag}) holds. Below this scale, the
coupling $\beta$ runs according to its two-dimensional renormalization-group
flow.

The sigma model (\ref{cp}) presents a nontrivial part of the worldsheet dynamics.
It is {\em not} supersymmetric.
Besides  (\ref{cp}), there are translational and supertranslational moduli;
they are represented by free bosonic and fermionic fields (two and four degrees
of freedom, respectively). Since these are free fields they
are not so important in what follows.

The sigma model (\ref{cp}) is asymptotically free \cite{Po3}; at large distances
(low energies) it gets into the strong coupling regime.  The  running
coupling constant  as a function of the energy scale $E$ at one loop is given by
\beq
4\pi \beta = N\ln {\left(\frac{E}{\Lambda_{{\rm CP}(N-1)}}\right)}
+\cdots,
\label{sigmacoup}
\eeq
where $\Lambda_{{\rm CP}(N-1)}$ is the dynamical scale of the CP$(N-1)$
model. As was mentioned above,
the UV cut-off of the sigma model at hand
is determined by  $g_{M2}\sqrt{\xi}$.
Hence,
\beq
\Lambda^N_{{\rm CP}(N-1)} = g_{M2}^N\, \xi^{N/2} \,\, \exp\left( -\frac{8\pi^2}{g_{M2}^2}\,\,I\right) \,.
\label{lambdasig}
\eeq
Note that in the bulk theory, due to the squark field VEVs, the coupling constant is frozen at
$g_{M2}\sqrt{\xi}$.

The coupling constant $g_{M2}$ is determined by the scale $\Lambda_{M,le}$
(see Eq. (\ref{Lambda})) of
the bulk monopole theory (\ref{mmodel}). Then  Eq.~(\ref{lambdasig}) implies
\beq
\Lambda_{{\rm CP}(N-1)}=\frac{\Lambda_{M,le}^{\,\,2I}}{(g_{M2}\sqrt{\xi})^{2I-1}}
\ll \Lambda_{M,le} \,,
\label{cpscale}
\eeq
where we take into account  that the first coefficient of the $\beta$ function
in (\ref{mmodel}) is $2N$.

Concluding this section let us add a few words on the fermion zero
modes on the non-Abelian string. We first note that the index theorem
presented in Ref.~\cite{Mmodel}
is valid only in the  $M_D$ model. It cannot be generalized to the $M_F$ model.
Therefore, we do not expect any superorientational fermion zero modes of the string. Of
course, four supertranslational fermion zero modes are guaranteed by
``non-BPSness"
of the string at hand. They are ``Goldstinos'' of the \none bulk supersymmetry broken
by the string solution. They decouple from the worldsheet
CP$(N-1)$ model.   We have mentioned this fact above.

\subsection{Higher derivative corrections}
\label{highder}

The
CP$(N-1)$ model (\ref{cp}) is an effective theory which describes the
non-Abelian string dynamics only at low energies. It has
higher derivative corrections which become important at higher
energies. Higher derivative corrections run in powers of
\beq
\Delta\, \pt_k,
\label{hder}
\eeq
where $\Delta$ is a string transverse size.

In the $M_F$ model the string size is
determined by the inverse mass of the bulk states,
$$
\Delta\sim \frac1{g_{M2}\sqrt{\xi}}\,.
$$
A typical energy scale on the string
worldsheet is   $\Lambda_{{\rm CP}(N-1)}$,
see (\ref{cpscale}). Thus,
$$\pt\to\Lambda_{{\rm CP}(N-1)}\,,$$
and higher derivative corrections can be ignored.

However, the monopole theory (\ref{mmodel})
is not quite the $M$ model. In addition to the field content of
the $M$ model we have more  light
$M$ fields in the bulk, namely the Goldstone ($M^P_S$) and
pseudo-Goldstone ($M^{\dot K}_P$, $M^P_{\dot L}$) states.
The fact of their
presence entails that the string profile functions
 acquire long-range tails at the quantum level \cite{SYnone}
(classically, as we have already mentioned, the fields $M^A_B$ vanish on the
string solution).
This means that an effective string thickness  grows, and
higher derivative corrections to the basic
CP$(N-1)$ model on the string worldsheet {\em might} become important.

Let us show that this does not happen. First note that the Goldstone
states ($M^P_S$) are singlets with respect to the SU$(N)_{C+F}$ global
symmetry.  This means that the {\em orientational} zero modes of the string have no
long-range tails associated with the $M^P_S$ fields and are perfectly
normalizable.

As for the pseudo-Goldstone states $M^{\dot K}_P$ and  $M^P_{\dot L}$,
they are not singlets with respect to SU$(N)_{C+F}$.
Therefore,
 long-range profile functions of $M^{\dot K}_P$ and  $M^P_{\dot L}$ can
acquire $n^l$-dependence at the quantum level. Then the higher
derivative corrections associated with the pseudo-Goldstone fields
will run in powers of
\beq
\Delta_{\rm pG}\, \frac{\sqrt{\xi}}{{\mathcal M}}\,\,\pt_k
\sim  \frac{\sqrt{\xi}\,\Lambda_{{\rm CP}(N-1)}}{m_{\rm pG}\, {\mathcal M}}
\label{hderpG}
\eeq
where we take into account that the coupling of the pseudo-Goldstone fields
to the classical profile functions of the string is suppressed, see
Eq.~(\ref{LEsup}). Since $$\xi/(m_{\rm pG}\mathcal M)\sim 1$$
we conclude that the higher-derivative
corrections remain to be negligible.

\section{Implications of strings in the monopole theory}
\label{impli}
\setcounter{equation}{0}

We begin from a few technical remarks.
The strings we found  at weak coupling in the  monopole theory
are in one-to-one correspondence with the vacua of the worldsheet
theory. In reviewing this correspondence we
 will be brief since our discussion will run in parallel
to that of Ref.~\cite{GSY05} which presents the issue in great detail.
The non-supersymmetric CP$(N-1)$  model was solved by Witten
in the large-$N$ limit \cite{W79}. Interpretation of Witten's results
in terms of non-Abelian strings in four dimensions can be found also in the review
paper \cite{SYrev}.

The model (\ref{cp}) can be understood as a
strong coupling limit of an U(1) gauge theory. The action has the form
\beq
S =\int d^2 x \left\{
 2\beta\,|\nabla_{k} n^{l}|^2 +\frac1{4e^2}F^2_{kp}
+2e^2 \beta^2(|n^{l}|^2 -1)^2
\right\}\,,
\label{nscpg}
\eeq
where $\nabla_{k} =\pt_k-iA_k$.
In the limit $e^2\to \infty$ the U(1) gauge field $A_k$  can be
eliminated via the (algebraic) equation of motion which leads to the theory
(\ref{cp}). Moreover, the condition (\ref{unitvec}) is implemented in
the limit $e^2\to \infty$.

The non-supersymmetric CP$(N-1)$  model is asymptotically free and develops
its own dynamical scale
$\Lambda_{{\rm CP}(N-1)}$.
Classically the  field $n^{l}$ can have arbitrary
direction; therefore, one might naively expect
spontaneous breaking of SU$(N)_{C+F}$ and
the occurrence of massless Goldstone modes.
This cannot happen
in two dimensions. Quantum effects restore the
full symmetry making the vacuum unique. Moreover, the condition
$|n^{l}|^2=1$ gets in effect relaxed. Due to strong coupling
we have more degrees of freedom than in the original Lagrangian,
namely all $N$ fields $n$ become dynamical and acquire
masses $\Lambda_{{\rm CP}(N-1)}$. They become $N$-plets of SU$(N)$.

\begin{figure}
\epsfxsize=6cm
%\centerline{\epsfbox{vacuama}}
\centerline{\epsfbox{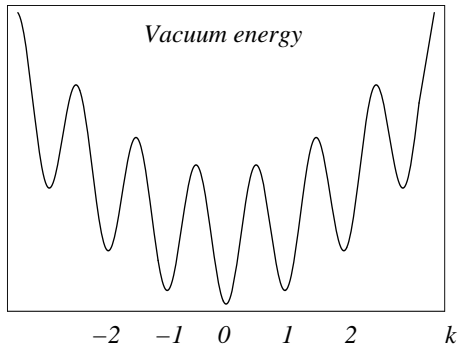}}
\caption{
The vacuum structure of
CP$(N-1)$ model.}
\label{odin}
\end{figure}

The modern understanding of the vacuum structure of the
CP$(N-1)$ model
\cite{Wtheta} (see also \cite{Stheta})
is as follows. At large $N$,
along with the unique ground state,
the model has $\sim N$ quasi-stable local minima, quasi-vacua,
which become absolutely stable at $N=\infty$,
see Fig.~\ref{odin}. The relative
splittings between the values of the energy density in the adjacent
minima
is of the order of $1/N$, while
the probability of the false vacuum decay is proportional to
$N^{-1}\exp (-N)$ \cite{Wtheta,Stheta}. The $n$
quanta are in fact $n$ kinks interpolating between the genuine vacuum and
the adjacent minimum. The spatial domain inside the $\bar{n} n$
meson is a ``bubble" of an excited quasi-vacuum state inside the
true vacuum --- that's why the $n$ kinks are confined along the string.

In the four-dimensional bulk
theory the above  vacua correspond to a variety of non-Abelian strings.
Classically all these strings have the same tension. Due to
quantum effects in the worldsheet  theory
the degeneracy is slightly lifted. Excited strings can in principle decay
into the ground-state string, but at large $N$ their lifetimes tend to infinity.

Now, let us ask ourselves: what is the physical meaning of these strings?

Non-Abelian strings are formed in the monopole theory
(\ref{mmodel}) upon the $h$-field condensation, see (\ref{qvev}).
The dual quark field $h$ represents monopoles of the quark theory.
Thus, from the standpoint of the original quark theory
the strings must be interpreted
as some flux tubes filled in by a chromoelectric field in a highly
quantum regime.
The string junctions of different elementary strings in the monopole
theory -- ``monopoles" of the monopole theory --
are seen as kinks $n$ in the effective theory on the string
worldsheet, see \cite{SYrev} for details.
They must act as
some quark-like objects in the original theory.
These objects transform according to the representation
 \underline{$N$} of unbroken global SU$(N)_{C+F}$.

\begin{figure}[h]
\epsfxsize=6cm
\centerline{\epsfbox{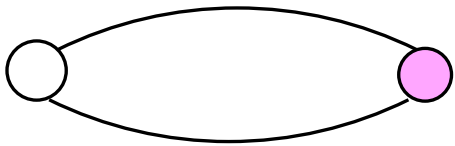}}
\caption{\footnotesize
The junction-antijunction meson. The binding is due to strings.}
\label{figmeson}
\end{figure}

The monopole theory strings can form closed curves (e.g. tori)
stabilized by angular momentum. They are to be interpreted
as sort of glueballs.
In addition, there are ``meson" states formed by junctions
connected by non-Abelian   strings,
see Fig.~\ref{figmeson}.
These mesons can belong either to the singlet or to the adjoint
representations of the global unbroken  SU(N)$_{C+F}$
symmetry. Both types of objects have masses $\sim \sqrt\xi$.

\section{Low-energy spectra and duality}
\label{conf}
\setcounter{equation}{0}

First let us summarize the low-energy spectrum of
the bulk theory as it is seen from the perspective of the monopole
theory.  The lowest are the Goldstone and pseudo-Goldstone states,
$M^P_S$ and $\{ M^{\dot K}_P,\,M^P_{\dot K}\}$, respectively. The latter
have masses
determined by $m_q=\xi/\mathcal M$, see (\ref{light}). From the
point of view of the quark theory these states can be understood as
quarks screened by the condensate of massless squarks. The Goldstone
states are singlets with respect to the unbroken SU$(N)_{C+F}$  while the
pseudo-Goldstone states transform in the  fundamental representation
of this group.

Next come states with masses of the order of $\sqrt{\xi}$. These set
includes elementary excitations:
gauge and $h$ multiplets as well as the fields $M_{\dot K}^{\dot L}$.
Their masses are determined by Eqs. (\ref{phmass}), (\ref{wmass}),
(\ref{U1mass}) and  (\ref{SUNmass}), see Sect.~\ref{spectrum}.
These states transform in the singlet or adjoint representations
of SU$(N)_{C+F}$. In addition, the set includes composite non-perturbative
states of the type we discussed at the end of Sect.~\ref{impli}.
The latter are believed to be metastable (rather than stable) as they can decay
into the massive gauge/monopole
multiplets  (with masses (\ref{phmass}) or (\ref{U1mass}))
with the appropriate quantum numbers with respect to the global
SU$(N)$.

In addition to
mesons in the low-energy part of the spectrum one can speak of
``baryons" built of  $N$ junctions cyclically connected to each other  by
elementary strings which form a closed ``necklace configuration."
(Here $N$ is not treated as an infinitely large parameter.
Of course, if $N\to \infty$, the baryons are out of the game.)
The baryon is in the
$\prod_1^N (N)$ representation of SU($N)_{C+F}$.

Note that both quarks and
monopoles do not carry the
baryon numbers. Therefore,  our ``baryon''  has no baryon number too.
The reason for this is that the U(1) baryon current is coupled to a gauge boson
(``photon") in  the U($N$) gauge
theory considered here. Moreover, the  U(1) gauge symmetry is spontaneously broken
in the quark and the monopole theories by condensation of quarks and dual quarks, respectively. Thus, the baryon charges are screened. This means, in particular, that  baryons can decay into mesons or gauge/monopole  multiplets and are in fact unstable.

All these nonperturbative states reflect the existence of ``thick"
strings with tension scaling as $\xi$ and thickness
proportional to $\xi^{-1/2}$.

All states with masses of the order of $\sqrt{\xi}$
will eventually decay into the Goldstone or pseudo-Goldstone mesons. Say,
a meson in Fig.~\ref{figmeson} in the
adjoint representation with respect to the global SU($N)$ can decay into a
pair of pseudo-Goldstone states. However, these decays are suppressed
by the smallness of the ratio $\sqrt{\xi}/{\mathcal M}$.

Now it is time to discuss the most interesting question:
if the low-energy states in the quark and monopole theories are connected by duality
how can one interpret the set of states we uncovered in the monopole theory
in the language of the quark theory?

The quark theory is strongly coupled. Quantitative predictions are virtually impossible.
Still we do have some qualitative knowledge of this theory.
In the quark theory color is screened since the theory is fully Higgsed.
There are matter fields in the fundamental representation.
Therefore long strings cannot exist. They are screened/ruptured immediately.
On the dual side we do see strings, however. The scale of the
string-induced confinement $\sqrt{\xi}$ is
small in the original quark theory, much smaller than its dynamical scale,
$\xi\ll\Lambda_Q^2$.

This apparent puzzle can be resolved if we assume that
a ``secondary" gauge theory (or a ``gauge cascade") develops
in the original quark theory. Assume that massless composite ``$\rho$ mesons"
whose size is $\sim \Lambda_Q^{-1}$ are formed
in the quark theory which interact with each other via a ``secondary" gauge theory
whose scale parameter is $\sqrt\xi$. At distances $\sim 1/\sqrt\xi$
the above ``$\rho$ mesons" must be viewed as massless gluons.
It is conceivable that they are coupled to
massless ``secondary" quarks which, in addition to their gauge coupling to
``$\rho$ mesons", have nontrivial quantum numbers with respect to
the global SU($N)$. With respect to the original quark theory
the ``secondary" quarks are colorless (``bleached") bound states which include
the original quarks at their core. Their sizes are proportional to $\sim \Lambda_Q^{-1}$
and, hence, they are pointlike on the scale of $\sim 1/\sqrt\xi$,
much in the same way as ``$\rho$ mesons"-gluons.

Alternatively one can adopt a more pragmatic
albeit less explicit point of view. If we trust
duality we can view the predictions derived in the monopole theory as
certain data to be interpreted in terms of the quark theory. One can think
that these predictions for the quark theory are similar to experimental
data for QCD.

Following this line of thought we can interpret the light spectrum seen from
the monopole theory as follows. First of all, we note that all states
in the physical spectrum of the
monopole theory are colorless. We interpret this as confinement
in the quark theory. Next, besides Goldstone and pseudo-Goldstone
states  (which we identify with $\tilde{Q}_S Q^P$ and
$\tilde{Q}_S Q^{\dot{K}}$ states)
the monopole theory predicts the occurrence of a set of light states
with masses of the order of $\sqrt{\xi}$. We interpret them in terms of
the quark theory according to their global flavor quantum numbers.
Say, we interpret singlets with respect to the global SU$(N)$   as
glueballs with possible admixture of quark-antiquark states
in the singlet representation. Next, we interpret perturbative and
non-perturbative states of the monopole theory in the adjoint
representation of SU$(N)$  as adjoint quark-antiquark mesons
or more complicated``exotic" multi-quark states.

Note, however, that we do not attempt to interpret
the monopole theory string junctions
as the fundamental quarks $Q^{\dot{K}}$ of the quark theory.
There is a number of
reasons why this identification does not work.
The monopole theory string junctions should be
rather understood as a kind of ``constituent quarks" which form mesons
in Fig.~\ref{figmeson}.  Although these ``constituent quarks" are in the
fundamental representation of the global flavor SU$(N)$ their relation to
the fundamental quarks $Q^{\dot{K}}$ of the quark theory Lagrangian
 remains unclear.

\section{Conclusions}

Our starting point is a Seiberg's dual pair with electric theory lying to the
left of the conformal window and the magnetic theory to the
right. The electric theory is strongly coupled while the magnetic one is
infrared free. Our basic idea is to deform the
electric theory very weakly --- with all deformations being very small in its
natural scale $\Lambda_Q$ --- and, nevertheless, they are sufficient to drastically
change the infrared behavior of the magnetic dual.
It switches from  infrared free to asymptotically free.
Seiberg's infrared duality now extends beyond purely massless states;
it connects with each other light states on both sides of duality.

Upon condensation of
the ``dual quarks," the dual theory supports
non-Abelian flux tubes (strings). Being interpreted in terms
of the quark theory these flux tubes are supposed to carry chromoelectric fields.
The string junctions then can be viewed as ``constituent quarks" of the original quark theory.
We interpret closed strings as glueballs of the original quark theory.
Moreover, there are
string configurations formed by two junctions connected by a pair of
different non-Abelian strings. These can be considered as constituent
quark mesons of the quark theory. Most of these states are quasistable rather than stable. They can cascade into the lightest Goldstones and pseudo-Goldstones.

The constituent quarks could result from emergent ``secondary gauge theory"
on the electric side.

\label{conclu}

 \section*{Acknowledgments}

We are grateful to N. Arkani-Hamed, A. Gorsky N. Seiberg and D. Shih for useful discussions.

The work of M.S. was
supported in part by DOE grant DE-FG02-94ER408.
The work of A.Y. was  supported
by  FTPI, University of Minnesota, by INTAS Grant No. 05-1000008-7865,
by RFBR Grant No. 06-02-16364a
and by Russian State Grant for
Scientific Schools RSGSS-11242003.2.

 \section*{Appendix A. Notation}

\renewcommand{\theequation}{A.\arabic{equation}}
\setcounter{equation}{0}

Flavor indices from SU($N_c+N$)
are denoted by capital letters from the beginning of the Latin alphabet,
\beq
A, B, ...= 1,2, ..., N_c, ..., N_c+N \,.
\eeq

\vspace{3mm}

Flavor indices from SU($N_c$)
are denoted by capital letters from the middle of the Latin alphabet,
\beq
P,S , ...= 1,2, ..., N_c \,.
\eeq

\vspace{3mm}

Flavor indices from SU($N_c+N$)/SU($N_c$)
are denoted by overdotted capital letters from the middle of the Latin alphabet,
\beq
\dot{K}, \dot{L} , ...= N_c+1,..., N_c+N \,.
\eeq

\vspace{3mm}

Color indices of the fundamental representation of
SU($N_c$) (and SU($N$) in the monopole theory)
are denoted by lower-case letters from the middle of the Latin alphabet,
\beq
k, l, ... = 1,2, .... N_c\,,\quad\mbox{or}\quad k, l, ... = 1,2, .... N\,.
\eeq

\vspace{3mm}

Color indices of the adjoint representation of
SU($N_c$) (and SU($N$) in the monopole theory)
are denoted by lower-case letters from the beginning of the Latin alphabet,
\beq
a,b, ... = 1,2, ..., N_c^2-1 \,,\quad\mbox{or}\quad a,b, ... = 1,2, ..., N^2-1\,.
\eeq

\vspace{13mm}

The bosonic part of the action of $M_F$ model is
\beqn
S&=&\int d^4x \left[\frac1{4\left( g_{M2}\right)^2}
\left(F^{a}_{\mu\nu}\right)^2 +
\frac1{4\left( g_{M1}\right)^2}\left(F_{\mu\nu}\right)^2+
\left|\nabla_{\mu}
h^{\dot K}\right|^2 + \left|\nabla_{\mu} \bar{\tilde{h}}^{\dot K}\right|^2
\right.
\nonumber\\[4mm]
&+& \frac2{\gamma} {\rm Tr}\,\left|\pt_{\mu} M\right|^2
+\frac{g^2_2}{2}
\left(
\bar{h}_{\dot K}\,T^a h^{\dot K} -
\tilde{h}_{\dot K} T^a\,\bar{\tilde{h}}^{\dot K}\right)^2
\nonumber\\[3mm]
&+& \frac{g^2_1}{8}
\left(
\bar{h}_{\dot K}\, h^{\dot K} -
\tilde{h}_{\dot K} \,\bar{\tilde{h}}^{\dot K}\right)^2+
 {\rm Tr}|hM|^2 +{\rm Tr}|\bar{\tilde{h}}M|^2
\nonumber\\[3mm]
&+&
\left.
\frac{\gamma}{2}\left|\tilde{h}_{\dot K} h^{\dot M} -\frac{1}{2}
\delta_{\dot K}^{\dot L}\delta_{\dot L}^{\dot M}\,\xi\right|^2
\right\}
\,.
\label{mmodelred}
\eeqn

\vspace{1mm}

\section*{Appendix B. The 't Hooft anomaly conditions}
%\addcontentsline{toc}{section}{Appendix A.}

\renewcommand{\theequation}{B.\arabic{equation}}
\setcounter{equation}{0}

That the quark and monopole theories are dual to each
other in the limit ${\mathcal Q}=m_q=0$
follows from Seiberg's construction.
Here we reiterate the analysis of the  't Hooft anomaly matching
relevant to our particular
deformation of the theory. There are certain peculiarities
since the matching looks different at ``high energies"
(i.e. when the momentum $q$ flowing in the axial curent
is ``large," $q^2\gg \xi$) and ``low energies" ($q^2\ll \xi$).
We check both limits.

As was already mentioned in Sect.~\ref{quarkth}   our theory has
the  following  global symmetry at the Lagrangian level
\beq
{\rm SU}(N_c+N)_L\times {\rm SU}(N_c+N)_R\times {\rm U}(1)_R\times
{\rm U}(1)_B
\label{glgrouptree1}
\eeq
where we also included the vector-like baryon U$(1)_B$ symmetry which is in
fact gauged in our model. The above  symmetry is non-anomalous with
respect to the non-Abelian gauge currents \cite{Sdual,IS}.  The fields of
the quark and monopole theories transform as
\beqn
&& Q \,:  \qquad
\left(N_c+N, 1, \frac{N}{N_c+N},\frac12\right)\,,
\nonumber\\[3mm]
&&
\tilde{Q} \,:  \qquad \left(1,N_c+N,  \frac{N}{N_c+N},-\frac12\right),
\nonumber\\[3mm]
&&
h \,: \qquad  \left(\bar{N}_c+\bar{N}, 1,
\frac{N_c}{N_c+N},\frac{N_c}{2N}\right),
\nonumber\\[3mm]
&&
\tilde{h} \,: \qquad  \left(1,N_c+N,
\frac{N_c}{N_c+N},-\frac{N_c}{2N}\right)
\label{treeQtrans}
\eeqn
under this symmetry, while the Grassmann
$\theta$ parameters have the unit charge under
$U(1)_R$ \cite{Sdual,IS}.

Consider first the quark theory.
Classically the tree-level  symmetry is broken down to
\beq
{\rm SU}(N)\times {\rm U}(1)_{R'}
\label{glQsym}
\eeq
by the condensation of the $Q^P$ fields and masses $m_q$ for $Q^{\dot K}$ fields.
Note that the $R$ symmetry U(1)$_{R'}$ classically survives the
breaking.  It is a combination of U(1)$_R$ and an axial subgroup of
non-Abelian factors in (\ref{glgrouptree1}) which do not transform
quark fields $Q^P$. It turns out that the fermion superpartners of
squarks $(\psi_Q)^{\dot K}$ and $(\tilde{\psi}_{Q})_{\dot K}$ have zero charges
with respect to this symmetry. Therefore, it is not broken by masses
$m_q$.

Quark fields of the quark theory transform
as
\beqn
&&
Q^{\dot K}
\,:
 \qquad \left(N, 1\right); \qquad
(\psi_Q)^{\dot K} \,:\qquad \left(N, 0\right),
\nonumber\\[3mm]
&&
Q^P
\,:
 \qquad \left(1, 0\right);  \qquad\,\,\,
(\psi_Q)^P \,:\qquad \left(1, -1\right),
\nonumber\\[3mm]
&&
\tilde{Q}_{\dot K}
\,:
\qquad \left(\bar{N}, 1\right); \qquad
(\tilde{\psi}_{Q})_{\dot K}\,: \qquad \left(\bar{N}, 0\right),
\nonumber\\[3mm]
&&
\tilde{Q}_P
\,:
\qquad \left(1, 0\right); \qquad\,\,\,\,
(\tilde{\psi}_{Q})_{P}\,: \qquad \left(1, -1\right)
\label{Qtrans}
\eeqn
under the classically unbroken  symmetry (\ref{glQsym}), while the
gauginos of the quark theory
($\lambda_Q$) transform as $(1,1)$.

In quantum theory, however, the U(1)$_{R'}$  is anomalous with respect
to the Abelian U(1) gauge currents (baryonic U(1)$_B$ currents).
 At high energies, well above the
scale of the U(1)$_B$ symmetry breaking,
the anomaly $U(1)_{R'}\, U(1)_B^2$ is proportional to
\beq
-2\left(\frac12\right)^2\, N_c^2=-\frac12\,N_c^2 ,
\label{qu1anom}
\eeq
which comes from
the contribution of $Q^P$ fermions. Here we take into account that we
have $N_c$ colors and  $N_c$  flavors of these fermions.  At low
energies the U(1)$_B$ charges are screened by the Higgs mechanism and the
anomaly effectively disappears.

Now consider the monopole theory. It has classically unbroken
\beq
{\rm SU}(N)_{C+F}\times {\rm U}(1)_{R''}
\label{glMsym}
\eeq
symmetry, where U(1)$_{R''}$ is a
combination of the original U(1)$_R$ and axial subgroup
of non-Abelian factors  in (\ref{glgrouptree1})   which do not
transform $h^{\dot K}$ and $\tilde{h}_{\dot K}$ fields.

Now, duality suggests us to identify two global classically unbroken groups
(\ref{glQsym}) and (\ref{glMsym}) of the quark and the monopole
theories.
The monopole fields $h$ transform as
\beqn
&&
h^{\dot K}
\,:
\qquad \left(\bar{N}, 0\right); \qquad
\psi^{\dot K}
\,:
\qquad \left(\bar{N}, -1\right),
\nonumber\\[3mm]
&&
h^P
\,:
\qquad \left(1, 1\right);  \qquad \,\,\,\,
\psi^P
\,:
\qquad \left(1, 0\right),
\nonumber\\[3mm]
&&
\tilde{h}_{\dot K}
\,:
\qquad \left(N, 0\right);
 \qquad
\tilde{\psi}_{\dot K} \,:
\qquad \left(N, -1\right),
\nonumber\\[3mm]
&&
\tilde{h}_P
\,:
\qquad \left(1, 1\right);
 \qquad \,\,
\tilde{\psi}_{P} \,:
\qquad \left(1, 0\right)
\label{Mtrans}
\eeqn
under the classically  unbroken  symmetry (\ref{glMsym}). The  $M$
fields of the monopole theory transform as
 \beqn
 && M^{\dot K}_{\dot L} \,:  \qquad
 \left(N^2, 2\right); \qquad (\psi_M)^{\dot K}_{\dot L}\,: \qquad \left(N^2,
1\right),
 \nonumber\\[3mm]
 && M^{\dot K}_P \,:  \qquad \left(N, 1\right);
 \qquad \,\,\, (\psi_M)^{\dot K}_P \,: \qquad \left(N, 0\right),
\nonumber\\[3mm]
&&
M_S^P
\,:
 \qquad \left(1, 0\right); \qquad \,\,\,\,
(\psi_M)^P_S
\,:
\qquad \left(1, -1\right)
\label{Mfieldtrans}
\eeqn
where $\psi_M$ are fermion superpartners of the scalar fields $M$.
The $\lambda$ fermions  transform as (1,1).
Note that the U(1)$_{R''}$ symmetry is not broken by the condensation
of the $M_S^P$ fields because these fields are neutral under this symmetry.

However, much in the same way as in the quark theory, the $R$ symmetry U(1)$_{R''}$
 is anomalous with respect to the U(1) gauge currents. At high energies,
above the scale of the U(1)$_B$ symmetry breaking, the anomaly
 $U(1)_{R''}\, U(1)_B^2$ is proportional to
\beq
-2\,\left(\frac{N_c}{2N}\right)^2\,N^2=-\frac12\, N_c^2,
\label{mu1anom}
\eeq
which comes from the contribution of $h^{\dot{K}}$ fermions.
Here we take into account that we have $N$ colors and $N$ flavors
of $h^{\dot{K}}$ fermions as well as their baryon charges, see
(\ref{treeQtrans}).
We see that the anomaly in the monopole theory matches with
the one in the quark theory (\ref{qu1anom}). At low energies the
U(1)$_B$ charges are screened and the anomaly effectively disappears.

Since the $R$ symmetry is classically
unbroken we can check the 't Hooft anomaly
matching conditions for the quark and monopole theories. This calculation
is quite similar to that reported in \cite{Sdual}. The first anomaly
to check is
\beq
{\rm SU}(N)^2\,\,{\rm U}(1)_R \,:\qquad 0=-N+N,
\eeq
where the left-hand side and the right-hand side at high energies
 are given by the quark and monopole theories, respectively. On the
monopole-theory side we take into account the contributions of the $h$ and
$M$ fermions. At low energies both $h$ and $M$ fermions become
massive and do not contribute to the anomaly. The matching condition is
trivially satisfied.

Next we check
\beq
{\rm U}(1)_R^3\,: \qquad (-1)^3 \,2N_c^2 +(N_c^2-1)= (-1)^3\, 2N^2
+(N^2-1)+N^2+(-1)^3\, N_c^2\,,
\eeq
where at high energies on the quark theory side we take into account
the contributions of the $Q^P$ fermions and $\lambda_Q$ fermions, while
four contributions on the monopole-theory side are associated with the
$h$ fermions, $\lambda$'s and $M^K_L$ and $M^P_S$ fermions,
respectively.

At low energies this anomaly matching becomes
\beq
{\rm U}(1)_R^3\,: \qquad (-1)^3 \,N_c^2 = (-1)^3\, N_c^2\,,
\eeq
where we take into account that on the quark theory side  half of
the $Q^P$ fermions\,\footnote{In the bosonic sector only $2N_c^2$ squarks
$Q^P$  (out of $4N_c^2$ squark fields) are massless, see (\ref{dimHQ}).
Similar reduction happens in the fermionic sector by supersymmetry.}
and all $\lambda_Q$ fermions become massive and do not
contribute to the anomaly, while on the monopole theory side
the only contribution comes from the massless $M^P_S$ fermions.

Finally, the last  anomaly matching to check is
\beq
{\rm U}(1)_R \,:\qquad -2N_c^2 +(N_c^2-1)=-2N^2 + (N^2-1) +N^2 -N_c^2\,,
\eeq
where at high energies the quark-theory contribution comes from the
$Q^P$ fermions and $\lambda_Q$'s, while the monopole-theory contribution
comes from the $h$ fermions, $\lambda$'s, and  $M^{\dot K}_{\dot L}$ and $M^P_S$
fermions, respectively. At low energies we have
\beq
{\rm U}(1)_R\,: \qquad -N_c^2 =  -N_c^2\,,
\eeq
where the contribution on the quark theory side comes from a half of the
$Q^P$ fermions (those which are massless), while the contribution on
the monopole theory side comes from the $M^P_S$ fermions.

We see that all anomalies match.

%\newpage

\end{document}